\documentclass[aps,onecolumn,prd,nofootinbib,superscriptaddress,preprintnumbers]{revtex4}

\usepackage{graphicx,floatflt}
\usepackage{amsmath,amsfonts,amssymb,amsthm}
\usepackage{epsfig}

\usepackage{verbatim}
\usepackage{psfrag}
\usepackage{bm}
\usepackage{bbm}

\usepackage{epsf}
\usepackage{graphics}

\newcommand{\dd}{\mathrm{d}} 
\newcommand{\Mpl}{M_\mathrm{P}} 

\newcommand{\rr}{\mathrm}
\newcommand{\be}{\begin{equation}}
\newcommand{\ee}{\end{equation}}

\begin{document}
\preprint{TTK-12-12}

\title{Slow Roll during the Waterfall Regime:
The Small Coupling Window for SUSY Hybrid Inflation}

\author{S\'ebastien Clesse} \email{s.clesse@damtp.cam.ac.uk}
\affiliation{DAMTP, Centre for Mathematical Sciences, Wilberforce
  Road, Cambridge CB3 0WA (United Kingdom)}

\author{Bj\"orn~Garbrecht}  \email{garbrecht@physik.rwth-aachen.de}
 \affiliation{Institut f\"ur Theoretische Teilchenphysik und Kosmologie, RWTH Aachen University, 52056 Aachen, Germany} 
  
  \date{\today}

\begin{abstract}
It has recently been pointed out that a substantial amount of e-folds
can occur during the waterfall regime of hybrid inflation. Moreover,
Kodama {\it et.al.} have derived analytic approximations for the 
trajectories of the inflaton and of the waterfall fields. Based
on these, we derive here the consequences for $F$- and $D$-term
SUSY hybrid inflation:
A substantial amount of e-folds may occur in the waterfall regime,
provided $\kappa\ll M^2/M_{\rm P}^2$, where $\kappa$ is the superpotential coupling, $M$ the scale of symmetry breaking and
$M_{\rm P}$ the reduced Planck mass. When this condition is amply 
fulfilled, a number of e-folds much larger than $N_{\rm e}\approx60$
can occur in the waterfall regime and the scalar spectral index is then given by the expression found by Kodama {\it et.al.}
$n_{\rm s}=1-4/N_{\rm e}$. This value may be increased up to unity,
if only about $N_{\rm e}$ e-folds occur during the waterfall regime,
such that the largest observable scale leaves the horizon close to
the critical point of hybrid inflation, what can be achieved
for $\kappa\approx 10^{-13}$ and $M\approx 5\times 10^{12}{\rm GeV}$
in $F$-term inflation. Imposing the normalization
of the power spectrum leads to a lower bound on the scale of symmetry
breaking.
\end{abstract}


\maketitle

\section{Introduction}

Among the large variety of inflation models (for a recent review, see e.g.~\cite{Mazumdar:2010sa}), the hybrid class~\cite{Linde:1993cn, Copeland:1994vg} is particularly promising.   In hybrid models, inflation is realized in a false vacuum, along a nearly flat valley of the scalar field potential.  It ends with a waterfall phase, triggered when the inflaton field reaches a critical value, from which the potential in the transverse direction develops a tachyonic instability, forcing the fields to reach one of the global minima of the potential.  The facts that inflation can be realized at sub-Planckian field values and that it is based on renormalizable operators only are two attractive features of hybrid models.  Moreover, compared to most small field models~\cite{Spalinski:2009rq}, the initial conditions for the fields do not require any extreme fine-tuning because the inflationary valley is an attractor that can be reached from initial values located outside the valley~\cite{Clesse:2009ur, Clesse:2009zd, Clesse:2008pf, Lazarides:1996rk}.   

In the usual description of hybrid models~\cite{Linde:1993cn, Copeland:1994vg,Battye:2010hg,Battye:2006pk, Martin:2006rs, Rocher:2004my,Jeannerot:2005mc,Rehman:2009wv,Dine:2011ws}, inflation is assumed to stop nearly instantaneously
with the onset of the waterfall phase, and the dynamics is restricted to an effective 1-field slow-roll model.  Under these assumptions, the original version of hybrid inflation exhibits a slightly blue power spectrum and is therefore usually considered as ruled out by observations.  Moreover, when the $Z_2$ symmetry of the potential is broken at the end of inflation, domain walls are formed with catastrophic consequences for cosmology.  This problem can be solved by considering a complex auxiliary field, so that the broken symmetry is $U(1)$, leading to the formation of cosmic strings.  In this case, the power spectrum can be in agreement with the CMB data~\cite{Bevis:2007gh,Rocher:2004my}.  The fast waterfall phase itself has been the object of recent attention, especially to determine the contribution of iso-curvature perturbations~\cite{Lyth:2012yp, Gong:2010zf, Fonseca:2010nk, Abolhasani:2010kr, Alabidi:2006wa, Lyth:2010ch} as well as the level of non-Gaussianities~\cite{Enqvist:2004bk,Barnaby:2006km,Barnaby:2006cq,Alabidi:2006hg,Bugaev:2011wy,Bugaev:2011qt}.  

However, it has been pointed out recently that inflation can continue during the waterfall stage
for much more than 60 e-folds~\cite{Clesse:2010iz}, so that the observable perturbation scales exit the Hubble radius during the waterfall.   In such a case, the power spectrum of scalar perturbations is red, possibly in agreement with CMB observations~\cite{Clesse:2010iz,Kodama:2011vs,Mulryne:2011ni,Abolhasani:2010kn,Avgoustidis:2011em}, and any topological defect formed at the critical point of instability is conveniently stretched outside the observable universe.  

Hybrid inflation can be embedded in various high energy frameworks, like Grand Unified Theories~\cite{Jeannerot:1997is,Jeannerot:2000sv, Fukuyama:2008dv}, string cosmology~\cite{Koyama:2003yc,Dvali:1998pa,Berkooz:2004yc, Davis:2008sa, Brax:2006yq, Kachru:2003sx}, extra-dimensions~\cite{Fairbairn:2003yx}, as well as supersymmetry (SUSY)~\cite{Dvali:1994ms, Copeland:1994vg, Binetruy:1996xj, Kallosh:2003ux, Clauwens:2007wc,Garbrecht:2006az} and supergravity~\cite{Halyo:1996pp,Binetruy:2004hh}.  
Supersymmetric
models are additionally attractive, because they offer
an explanation for the protection of the flatness of the inflationary
valley against radiative corrections.   $F$-term~\cite{ Dvali:1994ms,Copeland:1994vg} and $D$-term~\cite{Binetruy:1996xj, Halyo:1996pp} hybrid models are the most well known realization of hybrid inflation in supersymmetry.  In these models, radiative corrections lift up the flat directions of the potential, giving rise to a red power spectrum of scalar perturbations.
In the absence of additional non-renormalizable corrections, the
classic prediction for the scalar spectral index~\cite{Dvali:1994ms} is
$n_{\rm s}=0.98$.
These models have been studied intensively in the effective 1-field slow-roll approach.   When the contribution of cosmic strings formed at the end of inflation is taken into account, the primordial power spectrum has been shown to be in agreement (even if in tension because of large values of the spectral index) with CMB observations~\cite{Battye:2006pk,Battye:2010hg} in some regions of the model parameter space.
The predictions for the
spectral index can be lowered when non-renormalizable operators
are added to the potential~\cite{urRehman:2006hu,Garbrecht:2006az,Battye:2006pk,Battye:2010hg,Rehman:2009wv}, but the additional parameters limit
the predictivity of the models. Moreover, the resulting models
correspond to hilltop scenarios along the waterfall trajectory,
and the initial conditions for the scalar fields that lead to
phenomenologically viable inflation are less general than for the
original models.

However, the possibility that inflation can continue during the waterfall and affect the observable predictions has not yet been explored for the $F$-term and $D$-term models.  This is the main goal of this paper.  We use the method of Kodama et al.~\cite{Kodama:2011vs} to integrate the two-field slow-roll dynamics during the waterfall phase and identify that for the small coupling regime satisfying $ \kappa \ll M^2/M_{\rm P}^2$, where $\kappa$ is the superpotential coupling, $M$ the scale of symmetry breaking and
$M_{\rm P}$ the reduced Planck mass, inflation continues for more than 60-e-folds along waterfall trajectories. 

 In this case, the standard effective 1-field approach is not valid and the observable predictions are modified.  We evaluate the amplitude of the power spectrum of adiabatic perturbations, as well as its spectral index.  When the number of e-folds realized classically during the waterfall is much larger than $N_{\rr e}$, the number of e-folds between the time when observable modes leave the Hubble radius and the end of inflation, the spectral index is given by $n_{\rr s} = 1 - 4 / N_{\rr e} $.  We also calculate the amplitude of the power spectrum, and derive a new constraint on the scale of symmetry breaking. In the limit when the observable scales leave the Hubble radius near the critical instability point, the spectral index tends to unity, so that it is in principle to find model parameters that accommodate with any value of the spectral index in the rage $1 - 4 / N_{\rr e} < n_{\rr s} <1$.
 
In Section~\ref{sec:models}, the $F$- and $D$-term hybrid models are reviewed.  A common parametrization of their potential that is convenient in order to deal with their dynamics near the critical instability point is introduced.  The slow-roll parameters are derived and the slow-roll equations of motion are given.   In Section~\ref{sec:coarse}, we give a coarse picture of the waterfall dynamics that applies when
the number of e-folds in the waterfall regime is much larger
than 60 and calculate the amplitude and the spectral index of the  power spectrum of adiabatic perturbations.  Section~\ref{section:refined} is dedicated to a more precise analysis of the dynamics, and we show, that a spectral index close to unity
can arise, provided the largest scales, that are observable
today, left the horizon during inflation at the beginning of the
waterfall phase. In Section~\ref{sec:qudiff}, we present estimates
for the initial condition for the classical evolution
of the waterfall field, that is induced by quantum diffusion.
Section~\ref{sec:summary} contains a summary and conclusions.

\section{SUSY Hybrid Inflation Close to the Critical Point}  \label{sec:models}

\subsection{$F$-Term Inflation}

The superpotential for $F$-term inflation is given by~\cite{Dvali:1994ms,Copeland:1994vg}
\be
W=\kappa \widehat S(\widehat {\bar H} \widehat H-M^2)~,
\ee
where $\widehat S$ is a gauge singlet superfield and
$\widehat H$ ($\widehat{\bar H}$) are superfields
in the (anti-)fundamental representation of ${\rm SU}({\cal N})$.
This gives rise to tree-level scalar potential
\be
V_0=\kappa^2\left(|\bar HH-M^2|^2+|S\bar H|^2 + |S H|^2\right)\,,
\ee
where now $S$, $H$ and $ \bar H$ are complex scalar fields.
When $S$ acquires a vacuum expectation value (vev), while $\langle H \rangle = \langle \bar H \rangle = 0$ (angle brackets denote the vev), as it is the case along
the trajectory that supports hybrid inflation,
there are $\cal N$ Dirac fermions of mass $\kappa S$,
$\cal N$ complex scalar fields $H_+=\frac1{\sqrt{2}}(H+\bar H)$
of mass square $m^2_+=\kappa^2(|S|^2-M^2)$ and
$\cal N$ complex scalar fields $H_-=\frac1{\sqrt{2}}(H-\bar H)$
of mass square $m^2_-=\kappa^2(|S|^2+M^2)$. The canonically
normalized real scalar field $\sigma=\sqrt2 |S|$ is the inflaton
field, while $\psi=\sqrt2 H_+$ is the waterfall field. Note that
the $D$-term forces $|H|=|\bar H|$, implying that the
vev of $H_-$ is vanishing.

When $\langle H_- \rangle = 0$, the tree potential
is
\be
\label{V:Ftree}
V_0(\sigma,\psi)=\kappa^2M^4
\left[
\left(1-\frac{\psi^2}{4M^2}\right)^2
+\frac{\sigma^2\psi^2}{4M^4}
\right]=\frac{\kappa^2}{4}\sigma_{\rm c}^4
\left[
\left(1-\frac{\psi^2}{2\sigma_{\rm c}^2}\right)^2
+\frac{\sigma^2\psi^2}{\sigma_{\rm c}^4}
\right]~.
\ee
The degrees of freedom enumerated above
give rise to the one-loop corrections
\be
\label{V:rad}
V_1=\frac{\kappa^4{\cal N}}{128 \pi^2}
\Bigg\{
(\sigma^2-\sigma_{\rm c}^2)^2\log\left(\kappa^2\frac{\sigma^2-\sigma_{\rm c}^2}{2Q^2}\right)
+(\sigma^2+\sigma_{\rm c}^2)^2\log\left(\kappa^2\frac{\sigma^2+\sigma_{\rm c}^2}{2Q^2}\right)
-2\sigma^4\log\left(\kappa^2\frac{\sigma^2}{2Q^2}\right)
\Bigg\}~,
\ee
where $\sigma_{\rm c}=\sqrt{2} M$ is the critical value and $Q$ is an ultraviolet
cutoff. SUSY $F$-term hybrid inflation takes place in
the potential $V=V_0+V_1$. For $\sigma<\sigma_{\rm c}$,
the scalar fields are in the waterfall regime, which is the concern of the
present paper.

For the dynamics near the critical point $\sigma_{\rr c}$, the first derivative of the radiatively induced
potential,
\be
\label{DV:rad}
\frac{\partial V_1(\sigma)}{\partial \sigma}\Bigg|_{\sigma=\sigma_{\rm c}}
=\frac{\kappa^4{\cal N}}{8\pi^2}\sigma_{\rm c}^3\log 2\,,
\ee
is of importance.
The second derivatives are of order
$\kappa^4 M^2/(16\pi^2)\times {\cal O}(1)$. These
induce $\eta$-parameters $\eta=\kappa^2 (M_{\rm P}/\sigma_{\rm c})^2/(16\pi^2)\times {\cal O}(1)$, with $\Mpl$ the reduced Planck mass.
An $\eta$-parameter larger than one violates the slow-roll conditions. Therefore, inflation
is terminated due to the radiative corrections close to the critical point
provided $\kappa\gg(4\pi \sigma_{\rm c})/M_{\rm P}\times {\cal O}(1)$.  We show below that a substantial
amount of e-folds occurs after crossing the critical point provided the
stronger constraint $\kappa\ll \sigma_{\rm c}^2/M_{\rm P}^2$ holds.
Hence, neglecting the second derivatives
is a self-consistent approximation in that regime.
In the form of the effective potential~(\ref{V:rad}), the ${\cal O}(1)$ factor
would actually encompass a term that logarithmically diverges at the critical point,
due to the correction from the massless waterfall field. Moreover, when
$\sigma<\sigma_{\rm c}$, the potential~(\ref{V:rad}) is ill defined, because
of the negative mass-square instability of the waterfall field. This is due
to the limitations of the method of calculating the effective potential and does not 
indicate a singular behavior in the time-evolution of $\sigma$. One should expect
that the IR-divergence exhibited by the logarithm is regulated by the time-evolution
of sigma or perhaps the horizon size $H^{-1}$. However, even when the logarithm is large, in the regime
$\kappa\ll M^2/M_{\rm P}^2$, its coefficient is small enough such that we can neglect this effect that would be intricate to deal with 
theoretically. The corrections that lead to the first
derivative~(\ref{DV:rad}) of the effective potential originate from
fields with positive mass square around the critical point. Therefore
Eq.~(\ref{DV:rad}) reproduces the slope of the potential
at the critical point in a reliable manner.

\subsection{$D$-Term Inflation}

For this model, the superpotential is~\cite{Binetruy:1996xj, Halyo:1996pp}
\be
W=\kappa\widehat S \widehat{\bar H}\widehat H~,
\ee
and the $D$-term is
\be
D=\frac g2 \left(|H|^2-|\bar H|^2 +m_{\rm FI}^2\right)~.
\ee
The superfields $\widehat H$ and $\widehat{\bar H}$ are
in the one-dimensional representation of a ${\rm U}(1)$
gauge group, and $m_{\rm FI}$ is the Fayet-Iliopoulos
term. The canonically normalized inflaton field
is $\sigma=\sqrt 2|S|$ and the waterfall field $\psi=\sqrt 2 |\bar H|$.
For $\langle \psi \rangle=0$, there are two real scalar
fields of mass square $\kappa^2 \sigma^2/2+g^2m_{\rm FI}^2/4$
and two of mass square $\kappa^2 \sigma^2/2-g^2m_{\rm FI}^2/4$.
When the field $\sigma$ evolves below its critical value
\be
\sigma_{\rm c}=\frac1{\sqrt{2}}\frac g\kappa m_{\rm FI}\,,
\ee
the mass square of the waterfall field becomes negative.
In addition, there is a Dirac fermion of mass square $\kappa^2\sigma^2/2$.
This leads to the tree-level potential
\begin{align}
\label{V:Dtree}
V_0&=\kappa^2\left(
|H\bar H|^2+|SH|^2+|S\bar H|^2
\right)
+\frac12 D^2
\\\notag
&=\frac{g^2}{8}m_{\rm FI}^2
\left[
\left(1-\frac{\psi^2}{2m_{\rm FI}^2}\right)^2+2\frac{\kappa^2}{g^2}\frac{\sigma^2\psi^2}{m_{\rm FI}^4}
\right]
\\\notag
&=\frac{\kappa^4}{2 g^2}\sigma_{\rm c}^4
\left[
\left(
1-\frac{g^2}{4\kappa^2\sigma_{\rm c}^2}\psi^2
\right)^2
+\frac{g^2}{2\kappa^2\sigma_{\rm c}^4}\sigma^2\psi^2
\right]\,.
\end{align}
When eliminating $m_{\rm FI}$ in favor of $\sigma_{\rm c}$, the one-loop potential
takes the same form as for the $F$-term case, Eq.~(\ref{V:rad}). The $\eta$-term
at the critical point may be estimated as
$\eta=g^2 (M_{\rm P}/\sigma_{\rm c})^2/(8\pi^2)\times {\cal O}(1)$.
Therefore we must require that $g\ll2\sqrt{2}\sigma_{\rm c}/M_{\rm P}$, or,
equivalently $\kappa\ll2 m_{\rm FI}/M_{\rm P}$, for inflation not to
terminate at the critical point. The comments on the IR-divergence
of the second derivative and its relevance made above
for the $F$-term model apply to the present case as well.

\subsection{Common Parametrization}

Since the $F$- and $D$-term models share common features,
it is convenient to use the parametrization
\be
\label{V:general}
V=\Lambda\left[\left(1-\alpha \frac{\psi^2}{\sigma_{\rm c}^2}\right)^2 +2\alpha \frac{\sigma^2 \psi^2}{\sigma_{\rm c}^4}\right]
+\beta\sigma_{\rm c}^3\sigma\,.
\ee
The values of the particular parameters can be inferred from
the expressions~(\ref{V:Ftree},\ref{V:rad},\ref{V:Dtree}), and
they are summarized as well in Table~\ref{Table:pars}.

\begin{table}[ht!]
\begin{center}
\begin{tabular}{|c||c|c|}
\hline
 & $F$-term & $D$-term\\
\hline
$\Lambda$&$\kappa^2 M^4$&$\frac{\kappa^4}{2 g^2}\sigma_{\rm c}^4=\frac{g^2}{8} m_{\rm FI}^4$\\
\hline
$\sigma_{\rm c}$&$\sqrt 2 M$&$\frac{g}{\sqrt 2\kappa} m_{\rm FI}$\\
\hline
$\alpha$&$\frac12$&$\frac{g^2}{4\kappa^2}$\\
\hline
$\beta$&$\frac{{\cal N}\kappa^4}{16\pi^2}\log 2$&$\frac{\kappa^4}{16\pi^2}\log 2$\\
\hline
\end{tabular}
\end{center}

\caption{\label{Table:pars}
Parameters to be substituted into the
potential~(\ref{V:general}) in order to obtain the $F$- and $D$-term
models close to the critical point.}
\end{table}

We then follow Ref.~\cite{Kodama:2011vs} in introducing the parametrization
\begin{subequations}
\begin{align}
\sigma=&\sigma_{\rm c}{\rm e}^\xi\,,
\\
\psi=&\psi_0{\rm e}^\chi\,.
\end{align}
\end{subequations}
Throughout the slow roll-regime and after the crossing of the critical point,
$\xi<0$ and $|\xi|\ll1$,
which is consistently verified by the explicit solutions.
It is useful to note the derivatives
\begin{subequations}
\begin{align}
\label{DVDsigma}
\frac{\partial V}{\partial \sigma}=&
\Lambda \frac{4\alpha \sigma \psi^2}{\sigma_{\rm c}^4}
+\beta \sigma_{\rm c}^3\,,
\\
\label{DVDpsi}
\frac{\partial V}{\partial \psi}=&
\Lambda
\left(
\frac{4\alpha^2\psi^3}{\sigma_{\rm c}^4}
+\frac{4\alpha\psi}{\sigma_{\rm c}^2}
\frac{\sigma^2-\sigma_{\rm c}^2}{\sigma_{\rm c}^2}
\right)
\approx
\Lambda
\left(
\frac{4\alpha^2\psi^3}{\sigma_{\rm c}^4}
+\frac{8\alpha\psi\xi}{\sigma_{\rm c}^2}
\right)
\,,
\\
\frac{\partial^2 V}{\partial\sigma^2}
=&
\Lambda\frac{4\alpha\psi^2}{\sigma_{\rm c}^4}
\,,
\\
\frac{\partial^2 V}{\partial\psi^2}
=&\Lambda
\left(
\frac{12\alpha^2\psi^2}{\sigma_{\rm c}^4}
+\frac{4\alpha}{\sigma_{\rm c}^2}
\frac{\sigma^2-\sigma_{\rm c}^2}{\sigma_{\rm c}^2}
\right)
\approx
\Lambda
\left(
\frac{12\alpha^2\psi^2}{\sigma_{\rm c}^4}
+\frac{8\alpha\xi}{\sigma_{\rm c}^2}
\right)
\,,
\\
\frac{\partial^2 V}{\partial\sigma\partial\psi}
=&
\Lambda\frac{8\alpha\sigma\psi}{\sigma_{\rm c}^4}
\,
\end{align}

\end{subequations}
and the slow-roll parameters
\begin{subequations}
\begin{align}
\varepsilon_\sigma=&
\frac12M_{\rm P}^2
\left(
\frac{4\alpha \sigma \psi^2}{\sigma_{\rm c}^4}
\right)^2+\frac12 M_{\rm P}^2\frac{\beta^2\sigma_{\rm c}^6}{V_0^2}
\,,\\
\varepsilon_\psi=&
\frac12M_{\rm P}^2
\left(
\frac{4\alpha^2\psi^3}{\sigma_{\rm c}^4}
+\frac{4\alpha\psi}{\sigma_{\rm c}^2}
\frac{\sigma^2-\sigma_{\rm c}^2}{\sigma_{\rm c}^2}
\right)^2
\approx
\frac12M_{\rm P}^2
\left(
\frac{4\alpha^2\psi^3}{\sigma_{\rm c}^4}
+\frac{8\alpha\psi\xi}{\sigma_{\rm c}^2}
\right)^2
\,,\\
\eta_{\sigma\sigma}=&
M_{\rm P}^2\frac{4\alpha\psi^2}{\sigma_{\rm c}^4}
\,,\\
\eta_{\psi\psi}=&
M_{\rm P}^2
\left(
\frac{12\alpha^2\psi^2}{\sigma_{\rm c}^4}
+\frac{4\alpha}{\sigma_{\rm c}^2}
\frac{\sigma^2-\sigma_{\rm c}^2}{\sigma_{\rm c}^2}
\right)
\approx
M_{\rm P}^2
\left(
\frac{12\alpha^2\psi^2}{\sigma_{\rm c}^4}
+\frac{8\alpha\xi}{\sigma_{\rm c}^2}
\right)
\,,\\
\eta_{\sigma\psi}=&
M_{\rm P}^2\frac{8\alpha\sigma\psi}{\sigma_{\rm c}^4}
\,,
\end{align}
\end{subequations}
where $\varepsilon_X=\frac12M_{\rm P}^2[(\partial V/\partial X)/V]^2$
and $\eta_{XY}=M_{\rm P}^2[\partial^2V/(\partial X\partial Y)]/V$.
The first derivatives enter the slow-roll equations of motion,
\be
\label{eom:slowroll}
3H\frac{\partial X}{\partial t}=-\frac{\partial V}{\partial X}\,,
\ee
\be \label{eom:slowroll2}
H^2 = \frac{V}{3 \Mpl^2}~.
\ee

\section{Dynamics of the Waterfall: Coarse Picture} \label{sec:coarse}

\label{section:coarse}

In this Section, we determine the field trajectories for the last $N_{\rr e} $ e-folds of inflation that are relevant for CMB observations and calculate the scalar power spectrum amplitude and spectral index, in the generic case where inflation along the waterfall trajectories lasts for much more than $N_{\rr e} $ e-folds.   This regime corresponds to 
\be
\label{ass:coarse:1}
\alpha\psi^2/\sigma_{\rm c}^2\ll |\xi|
\ee
and
\be
\label{ass:coarse:2}
4\Lambda\alpha\psi^2\gg\beta\sigma_{\rm c}^6\,.
\ee
It is referred as phase~2(a) in Ref.~\cite{Kodama:2011vs}. 
Eqs.~(\ref{DVDsigma}) and~(\ref{DVDpsi}) and the slow-roll
equations of motion~(\ref{eom:slowroll}) yield
\be
\label{diffrel:phase2}
\frac{d\xi}{d\chi}=\frac12\frac{\psi_0^2}{\sigma_{\rm c}^2}\frac{{\rm e}^{2\chi}}{\chi}\,.
\ee
This relation can be integrated,
\be
\label{sol:coarse}
\xi^2=\frac 12 \frac{\psi_0^2}{\sigma_{\rm c}^2}{\rm e}^{2\chi}=\frac{\psi^2}{2\sigma_{\rm c}^2}\,.
\ee
A sufficient condition for inflation to terminate is the violation of the slow-roll condition
$|\eta_{\psi\psi}|<1$, that occurs for
\be
\label{xiend}
\xi=\xi_{\rm end}=-\frac{\sigma_{\rm c}^2}{8\alpha M_{\rm P}^2}~.
\ee
At this point, $\eta_{\sigma \psi}=\sqrt 2$, while
$\eta_{\sigma \sigma}= (\sigma_{\rm c}^2)/(8\alpha M_{\rm P}^2)\ll 1$.
Qualitatively, one may therefore explain the end of inflation as a consequence of
the classical backreaction via the dimensionless couplings of both evolving fields,
$\sigma-\sigma_{\rm c}$ and $\psi$, {\it cf.} also the
discussion in Ref.~\cite{Abolhasani:2010kr}.
Note that the $\eta$-conditions imply that
up to this point, no exponential tachyonic growth of modes of $\psi$ and
$\sigma$ has yet occured, such that it is justified to neglect quantum backreaction.

As a consistency check, we notice that the condition~(\ref{ass:coarse:1})
is met for $\xi=\xi_{\rm end}$, provided $\sigma_{\rm c}/M_{\rm P}\ll2$,
such that the vev of the inflaton during inflation is sub-Planckian and
the effects of (super-)gravity are perturbatively small.
At this point, we also find $\eta_{\sigma\psi}=1$ and
$\eta_{\sigma\sigma}=\sigma_{\rm c}^2/(8\alpha M_{\rm P}^2)$.
This latter term is much smaller than one in the $F$-term model, provided
$M\ll M_{\rm P}$ and in the $D$-term model provided $m_{\rm FI}\ll M_{\rm P}$.
These conditions coincide with those one would impose from demanding
(super-)gravity to be in the perturbative regime during inflation.
The initial conditions chosen for the particular solution~(\ref{sol:coarse})
correspond to a trajectory
that trespasses the critical point where $\xi=0$ and $\psi=0$.
At the critical point however, the assumptions~(\ref{ass:coarse:1},\ref{ass:coarse:2})
are not valid. Nevertheless, if there are values of $\xi$ such that
$\xi\ll \xi_{\rm end}$ and the
assumptions~(\ref{ass:coarse:1},\ref{ass:coarse:2}) hold on the
trajectory~(\ref{sol:coarse}), then Eq.~(\ref{sol:coarse})
corresponds to a trajectory evolving from somewhere {\it close} to the critical point
to $\xi_{\rm end}$.

Using the relation $dN=H dt$, where $N$ parametrizes the number of e-folds,
and substituting the trajectory~(\ref{sol:coarse}) into Eq.~(\ref{DVDsigma})
leads to the equation
\be
\label{soln:coarse}
\frac{d \xi}{d N}=-\frac{8M_{\rm P}^2 \alpha}{\sigma_{\rm c}^2}\xi^2\,,
\ee
which can be integrated to
\be \label{eq:phase1}
\xi=-\frac{\sigma_{\rm c}^2}{8 \alpha M_{\rm P}^2(N_{\rm end}-N+1)}\,.
\ee
How many e-folds can occur in this regime? The trajectory~(\ref{sol:coarse})
violates the condition~(\ref{ass:coarse:2}), when
$\xi=-\frac{\kappa}{4\pi}\sqrt{{\cal N}\log 2}$, where we set ${\cal N}=1$ for
the $D$-term case.

Therefore, in order to achieve more than $N_{\rm e}$ e-folds on the
trajectory~(\ref{sol:coarse}), the condition
\be
\label{rel:kappa}
\kappa\ll\frac{\pi\sigma_{\rm c}^2}{2\alpha M_{\rm P}^2\sqrt{{\cal N}\log 2}(N_{\rm e}+1)}
\ee
must be satisfied. This is one of our main results for supersymmetric hybrid
inflation in the waterfall regime. 

In turns out, as described in
Section~\ref{section:refined}, that a substantial amount of e-folds may also occur
before the violation of condition~(\ref{ass:coarse:2}).
In Ref.~\cite{Kodama:2011vs}, this is referred to as phase~1.
The condition
for this to happen has the same parametric dependence on $\alpha$, $\sigma_{\rm c}$
and $M_{\rm P}$ as for the relation~(\ref{rel:kappa}), but a different coefficient.
Before moving to that analysis, we derive the predictions for the power spectrum,
provided that inflation in the last $N_{\rm e}$ e-folds is described by the trajectory~(\ref{sol:coarse}), {\it i.e.} relation~(\ref{rel:kappa})
is satisfied. 

The instantaneous direction of the
evolution of the fields can be parametrized by
\be
\cos\vartheta=\frac{\dot\sigma}{\sqrt{\dot\sigma^2+\dot\psi^2}}\,,
\ee
where the dot denotes a derivative with respect to $t$. We refer to
the linear combination of fields in that direction by $s$, which
leads to the subscripts that we use in the following.
On the trajectory~(\ref{sol:coarse}), $\cos\vartheta=1/\sqrt3$.
The $\eta$ parameter in the $s$-direction is
\be
\label{eta:ss}
\eta_{ss}=\eta_{\sigma\sigma}\cos^2\vartheta+2\eta_{\sigma\psi}\cos\vartheta\sin\vartheta+\eta_{\psi\psi}\sin^2\vartheta\,.
\ee
The parameter $\eta_{\sigma\sigma}$ is suppressed by a factor of
$M^2/M_{\rm P}^2$ or $m_{\rm FI}^2/M_{\rm P}^2$, respectively.
The other parameters are evaluated for $N_{\rm end}-N=N_{\rm e}$,
using Eqs.~(\ref{soln:coarse}) and~(\ref{sol:coarse}).
For the scalar spectral index, one then finds~\cite{Kodama:2011vs}
\be
\label{ns:coarse}
n_{\rm s}=1+2\eta_{ss}=1-\frac4N_{\rm e}\,.
\ee
This result is generic for hybrid inflation in a wide parametric
range~\cite{Kodama:2011vs}, not only for the SUSY variants.
The number of e-folds $N_{\rm e}$ is to be evaluated at the fiducial
WMAP scale $k=0.002\,{\rm Mpc}^{-1}$.
In order to obtain the value for $N_{\rm e}$, we may approximate the energy scale of inflation 
that is dominated by $\Lambda$. We then obtain~\cite{Liddle:1993fq}
\begin{align}
\label{Ne}
N_{\rm e}=59.1- \frac{4-3(1+\bar w_{\rr{reh}})}{12(1+\bar w_{\rr{reh}})} \log\frac{\Lambda}{\varrho_{\rm reh}}+\frac 14 \log\frac{\Lambda}{(10^{16}\,{\rm GeV})^4}\,,
\end{align}
where $\varrho_{\rm reh}<\Lambda$ denotes the energy density, below which the Universe is radiation
dominated, and where $\bar w_{\rr{reh}} $ denotes the mean equation of state parameter during the reheating phase ({\it i.e.} the end of inflation and the point, after which the Universe is radiation-dominated).   Its value depends on the details of the tachyonic preheating phase.  If the tachyonic preheating process is not efficient, $\bar w_{\rr{reh}}  = 0$ and the Universe is matter-dominated
due to coherent oscillations of inflaton and waterfall field.

Provided that iso-curvature perturbations do not contribute significantly to the power spectrum, the value for $n_{\rm s}$ therefore lies somewhat below its most
recently reported central observational value~\cite{Komatsu:2010fb,Martin:2010kz}.
In order to maximize the value of $n_{\rm s}$, one should therefore assume $\varrho_{\rm reh}=\Lambda$, corresponding to instantaneous reheating. This can be achieved provided the fields $H$ and $\bar H$
have large couplings to the Standard Model sector. Nonetheless, as it stands, the
model is disfavored at more than 2$\sigma$ even if $N_{\rr e} \approx 60 $ by CMB observations.
However, note that
provided less than $N_{\rm e}$ folds lie on the
trajectory~(\ref{sol:coarse}), one may expect larger values for
$n_{\rm s}$, because the horizon exit then occurs for vevs at which
the curvature of the potential is smaller than in the present case.
We investigate this in Section~\ref{section:refined}.

Finally, we derive the additional constraint from the amplitude
of the power spectrum. The $\varepsilon$ parameter
in $s$-direction is
$\varepsilon_s=\varepsilon_\sigma+\varepsilon_\psi$.
For the amplitude, we obtain
\be
{\cal P}_{\cal R}=\frac{\Lambda}{24\pi^2 M_{\rm P}^4\varepsilon_s}
\Bigg|_{N_{\rm end}-N=N_{\rm e}}=
\frac{2^4\alpha^2\Lambda M_{\rm P}^2 N_{\rm e}^4} {9\pi^2 \sigma_{\rm c}^6}\,,
\ee
{\it i.e.}
\begin{subequations}
\label{P:R}
\begin{align}
\label{P:R:F}
{\cal P}_{\cal R}=&\frac{\kappa^2 M_{\rm P}^2 N_{\rm e}^4}{18 \pi^2 M^2}
\quad\textnormal{for $F$-term inflation}\,,
\\
\label{P:R:D}
{\cal P}_{\cal R}=&\frac{\kappa^4 M_{\rm P}^2 N_{\rm e}^4}{9 \pi^2 g^2 m_{\rm FI}^2}
\quad\textnormal{for $D$-term inflation}\,.
\end{align}
\end{subequations}
In conjunction with the constraint~(\ref{rel:kappa}), this leads
to lower bounds on the symmetry-breaking scales:
\begin{subequations}
\begin{align}
\frac{M^2}{M_{\rm P}^2}\gg\frac{9{\cal N}\log 2}{2 N_{\rm e}^2}{\cal P}_{\cal R}
\quad\textnormal{for $F$-term inflation}\,,
\\
\frac{m_{\rm FI}^6}{M_{\rm P}^6}\gg\frac{9 g^2 \log^2 2}{\pi^2}{\cal P}_{\cal R}
\quad\textnormal{for $D$-term inflation}\,.
\end{align}
\end{subequations}
These relations together with Eqs.~(\ref{P:R}) constitute another
main result for SUSY-hybrid inflation in the waterfall regime.

\begin{figure}[h!]
\begin{center}
\includegraphics[height=7cm]{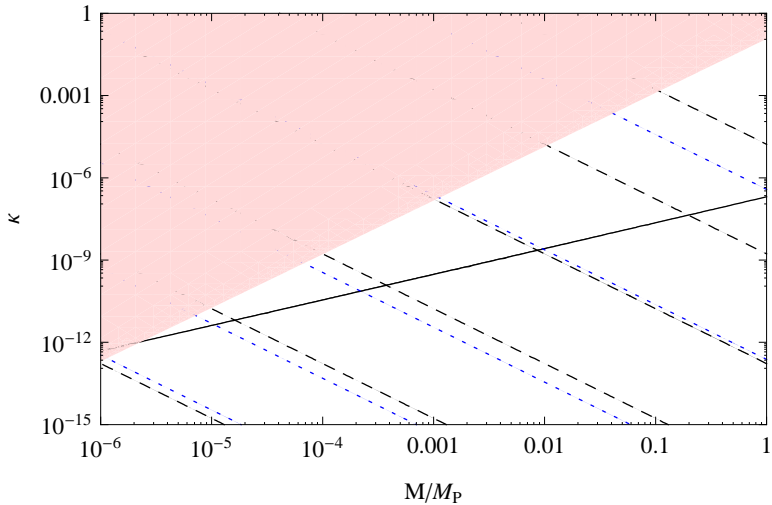}
\caption{
Relation between $\kappa$ and $M$ for the $F$-term model
with $\mathcal N =1$ (solid black line), that follows from the
amplitude of the power spectrum~(\ref{P:R:F}) and
the number of e-folds~(\ref{Ne}). The light red region corresponds to the regime of a nearly instantaneous waterfall transition, determined with Eq.~(\ref{rel:kappa}), for which no more than a few number of e-folds is realized along classical waterfall trajectories.  Dashed black lines are the iso-contours of constant energy scale for inflation ($V^{1/4}_*$), respectively $10^6$GeV, $10^8$GeV, $10^{10}$GeV, $10^{12}$GeV and $10^{14}$GeV from left to right.  Blue dotted lines are the iso-contours of constant spectral index from Eq.~(\ref{ns:coarse}), respectively $0.89, 0.90, 0.91, 0.92, 0.93$  from left to right.}
\label{fig:FtermCMB}
\end{center}
\end{figure}

In FIG.~\ref{fig:FtermCMB}, we plot the relation between
$\kappa$ and $M$ for the $F$-term model
with $\mathcal N = 1$ (notice that the influence of the parameter $\mathcal N$ is not very significant), that is
imposed by the normalization of the amplitude of
the power spectrum~(\ref{P:R:F}).
Moreover, we mark the region in which $\kappa$ is too large
(or $M$ is too small) in order to lead to a large enough amount
of e-folds~(\ref{Ne}).
Lower bounds on the mass parameter $M \gtrsim 10^{-6} \Mpl$ and on the coupling $\kappa \gtrsim 10^{-12}$ are deducted.
Finally, we have plotted the corresponding energy scale of inflation at the critical point of instability.  It is found to vary from $10^6$GeV to $10^{15}$GeV.

\begin{figure}[h!]
\begin{center}
\includegraphics[height=10cm]{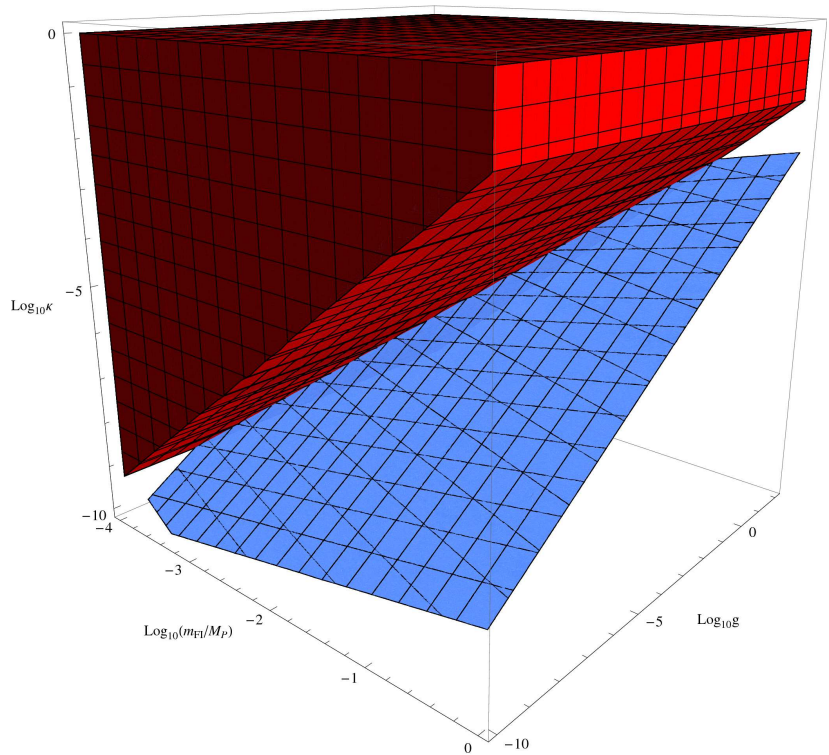}
\caption{Constraints on the 3D parameter space of $D$-term model from the normalization of the power spectrum~(\ref{P:R:D}) (blue area).
The red region corresponds to the regime of nearly instantaneous waterfall, determined with Eq.~(\ref{rel:kappa}).
\label{fig:DtermCMB}}
\end{center}
\end{figure}

\begin{figure}[h!]
\begin{center}
\includegraphics[height=7cm]{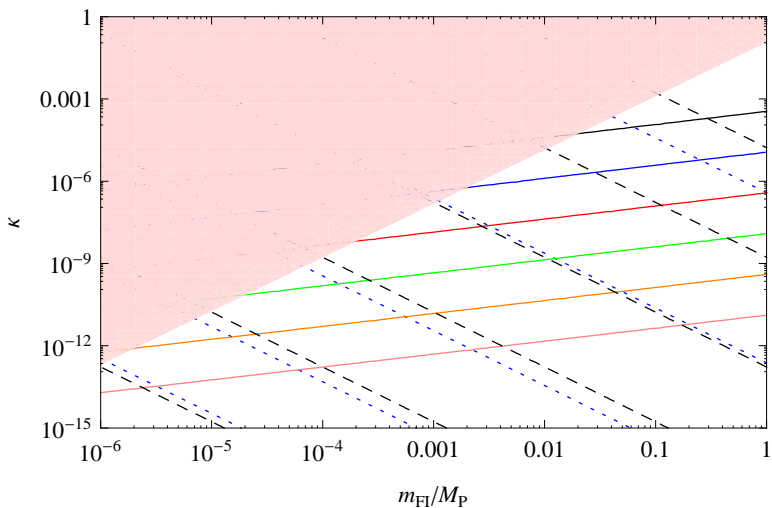}
\caption{
Relation between $\kappa$ and $m_{\rm FI}$ for various values
of $g$ (from top to bottom: $1$, $10^{-3}$, $10^{-6}$, $10^{-9}$, $10^{-12}$, $10^{-15}$) in the $D$-term model from the normalization of the power spectrum~(\ref{P:R:D}) (blue lines).
Above the red lines, that are determined by Eq.~(\ref{rel:kappa}), the waterfall transition proceeds nearly instantaneously. Black dashed lines and blue dotted lines are iso-contours of constant energy density and spectral index values, as in FIG~\ref{fig:FtermCMB}.
\label{fig:DtermCMB:2D}}
\end{center}
\end{figure}

CMB constraints on the three-dimensional parameter space of the $D$-term model are plotted in FIG.~\ref{fig:DtermCMB}.  The region allowed by the normalization of the power spectrum~(\ref{P:R:D})
corresponds to a thin slice of this parameter space. We restrict $\kappa<4\pi$, in order to keep the perturbation theory expansion
valid. A corresponding two-dimensional diagram for various
values of $g$ is provided in FIG~\ref{fig:DtermCMB:2D}.   
For $0.1 \Mpl \lesssim m_{\rr{FI}} \lesssim \Mpl$, the spectral index can be in agreement (but in strong tension) with the WMAP constraints, provided a coupling to fermions of the order of unity and $\kappa \approx 10^{-4}$. 

Let us notice also that the energy scale of inflation for the $D$-term model can be as low as a few TeV and the number of e-folds during the waterfall about $N_{\rr e}$, provided $g\approx10^{-20}$, $\kappa \approx10^{-16} $ and $m_{\rr{FI}} \approx 10^{-8} \Mpl$.   This extreme case is of particular interest since the D-term model can then provide a mechanism for the recent acceleration of the Universe expansion~\cite{Ringeval:2010hf}. 

\section{Dynamics of the Waterfall: Refined Picture}
\label{section:refined}

We apply now some more of the details that are derived in
Ref.~\cite{Kodama:2011vs} to SUSY hybrid inflation.
In particular, we consider initial conditions
that satisfy relation~(\ref{ass:coarse:1}) but that
turn around the bound~(\ref{ass:coarse:2}), such that
\be
\label{ass:phase1}
4\Lambda\alpha\psi^2\ll\beta\sigma_{\rm c}^6\,.
\ee
In Ref.~\cite{Kodama:2011vs}, this is referred to as phase~1(a).
Use of the slow-roll equations of motions~(\ref{eom:slowroll})
and Eqs.~(\ref{DVDsigma}) and~(\ref{DVDpsi}) leads to the differential
relation
\be
\frac{d\xi}{d \chi}
=\frac{\beta \sigma_{\rm c}^4}{8\alpha \Lambda\xi}\,,
\ee
what can be integrated to
\be
\label{trj:ph1}
\xi^2=\frac{\beta\sigma_{\rm c}^4}{4\alpha \Lambda}(\chi -\chi_1)+\xi_1^2\,.
\ee
Here, $\xi_1$ and $\chi_1$ should be determined by appropriate boundary
conditions. The point on this trajectory, where
the condition~(\ref{ass:phase1}) is violated, is denoted
by $\xi_2$ and $\chi_2$. When $\xi_2\gg \xi_1$ and $\chi_2\gg \chi_1$,
one may approximate $\xi_1\approx 0$, $\chi_1\approx 0$, which
is what we assume in the following. Using these approximations,
we obtain
\begin{subequations}
\begin{align}
\label{chi2}
\chi_2=&\frac12\log \frac{\beta \sigma_{\rm c}^6}{4\alpha \Lambda \psi_0^2}\,,
\\
\xi_2^2=&\frac{\beta\sigma_{\rm c}^4}{4\alpha \Lambda}\chi_2
\,.
\end{align}
\end{subequations}

The number of e-folds before reaching $(\xi_2,\chi_2)$ is
\begin{subequations}
\begin{align}
N_2=&\frac{1}{2M_{\rm P}^2}\sqrt{\frac{\Lambda\chi_2}{\alpha\beta}}\,,
\\
N_2=&\frac{M^2}{\kappa M_{\rm P}^2}2\pi\sqrt{\frac{2\chi_2}{\log2}}\quad\textnormal{for $F$-term inflation}
\,,
\\
\label{N2D}
N_2=&\frac{m_{\rm FI}^2}{\kappa M_{\rm P}^2}2\pi\sqrt{\frac{\chi_2}{\log2}}\quad\textnormal{for $D$-term inflation}\,.
\end{align}
\end{subequations}

Therefore, also in the regime where relation~(\ref{ass:phase1})
is valid, a substantial amount of e-folds may occur in
the waterfall regime provided $\kappa\ll M^2/M_{\rm P}^2$
or $\kappa\ll m_{\rm FI}^2/M_{\rm P}^2$, respectively.

Substituting the trajectory~(\ref{trj:ph1}) into
Eqs.~(\ref{DVDsigma},\ref{eom:slowroll}) and making use
of the relation~(\ref{ass:phase1}), one finds
\be
\label{sol:phase1}
\xi(N)=-N\frac{M_{P}^2}{\Lambda}\beta\sigma_{\rm c}^2\,,
\ee
such that the fields are close to the critical point for $N=0$.

We check whether before reaching $\xi_2$, the
condition~(\ref{ass:coarse:1}) may be violated.
The inequality~(\ref{ass:coarse:1}) holds for all points
on the trajectory~(\ref{trj:ph1}) before reaching
$(\xi_2,\chi_2)$, provided that
\begin{subequations}
\begin{align}
\chi_2\gg&\frac{\alpha\beta\sigma_{\rm c}^4}{\Lambda}\,,
\\
\chi_2\gg&\frac{\kappa^2}{128\pi^2}\log2
\quad\textnormal{for $F$-term inflation}\,,
\\
\label{notempminD}
\chi_2\gg&\frac{g^4}{256\pi^2\kappa^2}\log2
\quad\textnormal{for $D$-term inflation}
\,.
\end{align}
\end{subequations}

Since $\chi_2$ is given the by logarithm in Eq.~(\ref{chi2}) and
$\kappa\ll 1$, it is immediately clear that above condition
holds for the $F$-term case (barring the choice of large values
for $\psi_0$). For $D$-term inflation, Eq.~(\ref{N2D}) and
relation~(\ref{notempminD}) combine to
\be
\frac{\kappa}{g}\gg\frac{m_{\rm FI}}{\sqrt{8N_2}M_{\rm P}}\,,
\ee
which is satisfied because we already observe the stronger
constraint
$\sigma_{\rm c}\ll M_{\rm P}\Leftrightarrow\kappa/g
\gg m_{\rm FI}/(\sqrt2M_{\rm P})$. In conclusion,
the condition~(\ref{ass:coarse:1}) is fulfilled for
both, $F$- and $D$-term inflation at all times during
phase~1(a), and when combining this with the results of
Section~\ref{section:coarse}, it is
fulfilled at all times during the slow-roll
regime as well.

After the transition from phase~1 to phase~2
[the point $(\xi_2,\chi_2)$],
condition~(\ref{ass:phase1}) no longer holds and is replaced
by~(\ref{ass:coarse:2}).
The fields $\xi$ and $\chi$ satisfy the differential
relation~(\ref{diffrel:phase2}).

Compared to the trajectory~(\ref{sol:coarse}), a solution
can be determined that takes accurate account of the boundary
conditions that arise at the end of phase~1~\cite{Kodama:2011vs}:
\begin{subequations}
\begin{align}
\label{trajectory:refined}
\xi(N)=&\frac
{
-(c^\prime -c) f(N) + c^\prime +c
}
{
(c^\prime-c)f(N)+c^\prime +c
}
\xi_2^\prime\,,
\\
f(N)=&{\rm e}^{16c^\prime M_{\rm P}^2\sqrt{\frac{\alpha\beta}{2\Lambda}}(N-N_2)}\,,
\\
c=&\sqrt{\chi_2/2}\,,
\\
c^\prime=&\sqrt{c^2-\frac14}\,,
\\
\xi_2^\prime=&-c^\prime\sigma_{\rm c}^2\sqrt{\frac{\beta}{2\alpha \Lambda}}\,.
\end{align}
\end{subequations}

At late times, the solution~(\ref{trajectory:refined}) approaches
the approximate form~(\ref{soln:coarse}). Notice that the
initial conditions $(\xi_2,\chi_2)$ depend on the initial condition
for phase~1 through $\psi_0$, while the late-time
behavior~(\ref{soln:coarse}) is independent of these.

We can substitute $\xi_{\rm end}$, Eq.~(\ref{xiend}), in
the solution~(\ref{trajectory:refined}), invert it and
obtain the number of e-folds in phase 2~\cite{Kodama:2011vs}:
\be
\label{e-folds:ph2}
N_{\rm end}-N_2
=\frac{1}{16c^\prime M_{\rm P}^2}\sqrt{\frac{2\Lambda}{\alpha\beta}}
\log
\left(
\frac{\xi_{\rm end}-\xi_2^\prime}{\xi_{\rm end}+\xi_2^\prime}
\frac{c+c^\prime}{c-c^\prime}
\right)\,,
\ee
which corresponds to an improved version of the estimate~(\ref{rel:kappa}).

\begin{figure}[h!]
\begin{center}
\includegraphics[height=6cm]{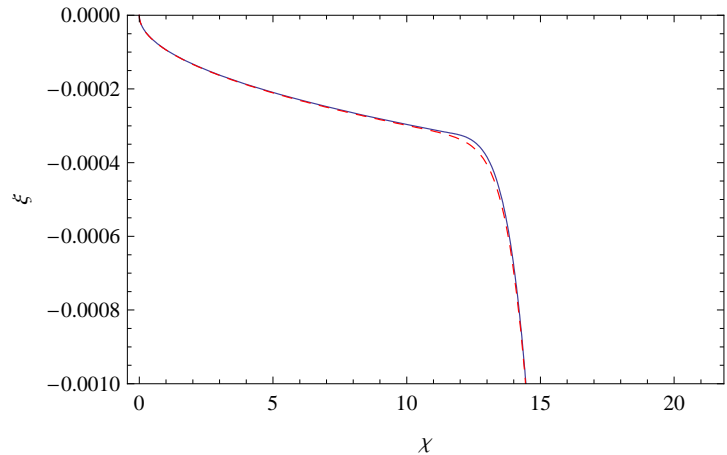}
\caption{Field trajectory in the $(\chi,\xi)$ space for F-term inflation, with $\kappa = 0.001$ and $M = 0.1 \Mpl$.  The plain blue curve is obtained from Eqs.~(\ref{eq:phase1},\ref{sol:coarse})   It is in agreement with the red dashed curve, obtained by integrating numerically the exact classical dynamics.}
\label{fig:Ftermtraj}
\end{center}
\end{figure}
It would be straightforward now to derive a general expression for
the power spectrum
in a situation where $N_{\rm end}-N_2<N_{\rm e}$, such that
the horizon exit of the largest observable scales occurs in phase~1(a).
The result is however somewhat complicated and shows no advantage
over a numerical evaluation, while having the disadvantage
of incurring an inaccuracy due to the matching of phases~1(a) and~2(a).
There is however an interesting limiting case, where simple estimates
can be obtained: While in Section~\ref{section:coarse}, we have
discussed the situation where the number of e-folds during the 
waterfall regime is much larger than $N_{\rm e}$, {\it i.e.}
$N_{\rm end}\gg N_{\rm e}$, we now consider
the situation where just enough e-folds occur in the waterfall regime,
{\it i.e.} $N_{\rm end}> N_{\rm e}$ and
$N_{\rm end}-N_{\rm e}\ll N_{\rm e}$.
When $\varepsilon_\sigma\gg\varepsilon_\psi$, as it is
the case sufficiently close to the critical point,
the amplitude of the power spectrum in phase~1(a) is
\be
\label{P:R:ph1}
{\cal P}_{\cal R}=\frac{\Lambda}{24\pi^2M_{\rm P}^2\varepsilon_\sigma}
=\frac{\Lambda^3}{12\pi^2\beta^6 M_{\rm P}^6\sigma_{\rm c}^6}\,.
\ee
The prediction for the scalar spectral index now depends
on the initial condition for $\psi$, {\it i.e.} the
choice of $\psi_0$ in the present parametrization.   
However, as explained in Sec.~\ref{sec:qudiff} below, the auxiliary field dynamics at the critical instability point is not classical but dominated by quantum diffusion effects.  The classical regime is nevertheless reached quickly and our initial value $\psi_0$ must be seen as the value that would take the auxiliary field if the classical trajectory was traced back up to the critical instability point.  

Note moreover that when $\dot \sigma\gg\dot \psi$,
\be
\cos\vartheta\approx1-N^2M_{\rm P}^4\frac{32\alpha^2\psi^2}{\sigma_{\rm c}^6}\,.
\ee
When we impose that the exit scale is close to the critical point,
the superpotential coupling must be of order
$\kappa\sim M^2/M_{\rm P}^2$ or $\kappa\sim m_{\rm FI}^2/M_{\rm P}^2$.
Close to the critical point, when assuming that $\psi$
is given by a value close to its
lower bound~(\ref{lowerbound:psi}), it then turns
out that the effective $\eta$-parameter~(\ref{eta:ss}) is
suppressed by $M^2/M_{\rm P}^2$ or $m_{\rm FI}^2/M_{\rm P}^2$. Therefore, the scalar spectral
index approaches values very close to unity when the horizon exit
of the largest observable scale occurs close to the critical point
of SUSY hybrid inflation.
It should therefore be possible to find parameters, that produce
all possible values for the scalar spectral index $n_{\rm s}$
between the values~(\ref{ns:coarse}) and {\it one}.
The parameters for which the horizon exit occurs close to
the critical point can be estimated from FIGS.~\ref{fig:FtermCMB},~\ref{fig:DtermCMB} and~\ref{fig:DtermCMB:2D} from the intersection of
the exclusion region, for which less than $N_{\rm e}$ e-folds
occur below the critical point
and the relation between $\kappa$ and $M$
or $m_{\rm FI}$, respectively. In particular, for $F$-term inflation,
we can estimate that $\kappa\approx 10^{-13}$
and $M\approx10^{12}\,{\rm GeV}$, in order to yield a value of
$n_{\rm s}$ close to its presently observed central value. A precise
determination of this point in parameter space and a quantitative
analysis of how much tuning is required, such that the spectral index
falls
within the allowed range, will be subject of a future study.

Calculating the possible values of $\kappa$, $M$ or $m_{\rm FI}$
for a given $n_{\rm s}$ using the present methods does not appear to
be possible in a simple analytic way. For example,
Eq.~(\ref{e-folds:ph2}) cannot be solved for $\kappa$ in terms
of elementary functions. Due to the simple nature of the slow-roll
equations, a numerical study should however be feasible.

\section{From quantum diffusion to classical dynamics} \label{sec:qudiff}

The statistical distribution of the initial auxiliary field values $\psi_0$ can be evaluated by studying the quantum diffusion near the critical instability point~\cite{Clesse:2010iz,Martin:2011ib}.  
The coarse-grained auxiliary field can be described by a Klein-Gordon equation to which a random noise field $\lambda (t)$ is added~\cite{Vilenkin:1983xp}.  This term acts as a classical stochastic source term.  In the slow-roll approximation, the evolution is given by the first order Langevin equation
\begin{equation}
\dot \psi + \frac{1}{3 H} \frac{\dd V}{\dd \psi}= \frac{H^{3/2}}{2 \pi} \lambda(t) ~,
\end{equation}
which can be rewritten by using Eq.~(\ref{DVDpsi})
\begin{equation} \label{eq:langevin}
\dot \psi \simeq \frac{H^{3/2}}{2 \pi} \lambda(t) + H \frac{4 \alpha \psi \Mpl^2}{\sigma_{\rr c}^2} \left( 1- \frac{\sigma^2}{\sigma_{\rr c} ^2}\right) ~.
\end{equation}
The two-point correlation function of the noise field obeys 
\begin{equation}   
\langle \lambda(t) \rangle = 0,  \hspace{5mm} \langle \lambda(t) \lambda(t') \rangle = \delta (t-t') ~.
\end{equation}
In the limit of $H$ constant (this approximation is valid at the critical instability point when the expansion is governed by the evolution of $\sigma$ in the false vacuum), this equation can be integrated exactly.  Under a convenient change of variable \cite{GarciaBellido:1996qt}, 
\be
x\equiv \frac{\sigma^2 }{ \sigma_{\rr c}^2} = \exp \left(-2 N\frac{M_{P}^2}{\Lambda}\beta\sigma_{\rm c}^2 \right)~,
\ee
 one gets
 \begin{equation}
\frac{\dd \psi}{\dd x} = - \frac{H^{1/2} } { 4 \pi r x} \lambda(x) - \frac {4 \alpha \psi \Mpl^2 (1-x)}{2 \sigma_{\rr c}^2 r x}~,
\end{equation}
with $r \equiv \Mpl^2 \beta \sigma_{\rr c} ^2 / \Lambda  $
This equation has an exact solution
\begin{equation}
\begin{aligned}
\psi(x) & =  C \exp \left( C_2 x - C_2 \ln x  \right) \\ 
 & - C_1 \exp \left( C_2 x - C_2 \ln x \right)   \times \int_1^x \exp \left( -C_2 x' + C_2 \ln x' \right) \lambda(x') \dd x' ~,
\end{aligned}
\end{equation}
where $C_1 \equiv H^{1/2} / ( 4 \pi r)  $, $C_2 \equiv 2 \alpha /  (\sigma_{\rr c}^2 r)   $ and $C$ is a constant of integration.  The variance of the auxiliary field distribution is then obtained by taking the two point correlation function of $\psi(x)$.  By assuming an initial delta distribution for $\psi$ at $\sigma \gg \sigma_{\rr c}$, one obtains 
\begin{equation} \label{eq:psi_qudist}
\langle \psi^2(x) \rangle = \frac {H^2}{8 \pi^2 r} \left[ \frac{\exp (x)}{ a x}  \right]^a  \Gamma(a,a x) \ ,
\end{equation}
where we have defined $a\equiv 4 \alpha \Mpl^2 /(\sigma_{\rr c}^2 r) = 4 \alpha \Lambda / (\beta \sigma_{\rr c}^4)$ and where  $\Gamma$ is the upper incomplete gamma function.
Near the instability, $x \simeq 1$ and one thus has
\begin{equation}  \label{eq:qufluct}
\langle \psi^2(x \simeq 1) \rangle \simeq \frac {H^2}{8 \pi^2 r} \left( \frac{\rr e}{a}  \right)^{a}  \Gamma \left(a , a  \right).
\end{equation}
By using recurrence relations as well as the asymptotic behavior of the $\Gamma$ function, one can find
\begin{equation}
\left(  \frac{\rr e}{u} \right)^u \times \Gamma(u,u) \sim \sqrt{\frac \pi 2} \frac 1 {\sqrt u} \hspace{2mm} \rr{when} \hspace{2mm} u \rightarrow \infty ~,
\end{equation} 
such that 
\begin{equation} \label{eq:variance_at_inst}
\langle \psi^2(x\simeq 1) \rangle \simeq \frac{H^2 }{8 \pi^{3/2} r \sqrt{2a}} = \frac{H^2 \sqrt{\Lambda}}{16 \pi^2 \Mpl^2 \sqrt{2 \pi \beta \alpha }}~. 
\end{equation}
At the critical instability point, the average value of $\psi $ over the whole Universe is zero, and Eq.~(\ref{eq:variance_at_inst}) describes the statistical distribution of the field around zero.   However, over a small patch that will contain our observable Universe, the average value is non zero and increases statistically with time due to the second term of Eq.~(\ref{eq:langevin}).   But the variance $\langle \psi^2\rangle$ in this patch is still given by Eq.~(\ref{eq:variance_at_inst}).  After some e-folds of inflation, the classical regime is reached and the classical evolution of $\psi$ proceeds faster than
quantum diffusion, $H^{-1}\dot\psi>H$.   By using
Eqs.~(\ref{DVDpsi},\ref{eom:slowroll},\ref{ass:phase1},\ref{sol:phase1}), one finds that this happens when
\be
\label{lowerbound:psi}
\psi>\frac{\Lambda^{3/2}}{8\sqrt 3 N \beta M_{\rm P}^5}~.
\ee
We should therefore see the classical dynamics of $\psi$ during the waterfall as emerging in a patch of the Universe where the quantum diffusion was previously dominating.  But the time when inflation takes place, the dynamics of the field $\sigma$ remains classical. In particular, when combining the estimates~(\ref{eq:variance_at_inst}) and~(\ref{lowerbound:psi}), one may see that typically, $N\ll 1$ when
the classical evolution begins: for $F$-term inflation,
$N\sim \kappa M^6/{\Mpl}^6$ and for $D$-term inflation,
$N\sim g m_{\rm FI}^6/{\Mpl}^6$.

Besides the problem of the quantum diffusion of the auxiliary field, one must also take care that the inflaton itself is classical.   For the original hybrid model, the regime dominated by the quantum stochastic fluctuations of the inflaton has been studied in Ref.~\cite{Martin:2011ib} and leads to a strong reduction of the number of e-folds realized during the waterfall.
This argument imposes the additional condition $|\dd \sigma / \dd N | \gg H / (2 \pi) $.  During the phase 1(a), the classical evolution of $\sigma$ is governed by (\ref{sol:phase1}), so that this condition can be rewritten 
\be
\frac{12 \pi^2 \Mpl^6 \beta^2 \sigma_{\rr c} ^6 }{\Lambda^3} \gg 1~.
\ee
It is satisfied provided $\kappa \Mpl^3 / M^3 \gg 1$ for the $F$-term model, and $\kappa \Mpl^3/ m_{\rr{FI}}^3 \gg 1$ for the $D$-term model.    It therefore appears for the $D$-term model that in the range  $0.1 \Mpl \lesssim m_{\rr{FI}} \lesssim \Mpl$, $g\sim \mathcal O(1)$ and $\kappa \approx 10^{-4}$ that was found to be in agreement with CMB observations, the quantum effects of $\sigma $ during the phase 1a can be important.  However, in that particular case, since the inflaton is driven by the second term of (\ref{DVDsigma}) during the last 60 e-folds of inflation (phase 2a), we argue that its dynamics is classical during this phase such that observable predictions are not affected by the quantum stochastic effects of $\sigma $ at the critical instability point. 

\section{Summary and Conclusions}
\label{sec:summary}

In the present work, we have explored the prospects of SUSY $F$-
and $D$-term models of accounting for the observed normalization
and spectral index of the primordial perturbation power spectrum.
We have focused on
the parametric regime where all scales that are observable today
have left the horizon during the waterfall stage. It is particularly
interesting to confront these scenarios with observational data,
because they only rely on the scale of symmetry  breaking and the
superpotential coupling for $F$-term inflation and, in addition,
the gauge coupling for $D$-term inflation. When the scale of symmetry
breaking is small compared to the Planck scale, the influence of
non-renormalizable operators, that is expected within the supergravity 
completion of these models, is suppressed. Besides, the trajectories
of the scalar fields are attracted to the valley of the hybrid
inflation potential, what makes these models rather predictive.

In order to derive our results, we use the analytical methods
that have been introduced in Ref.~\cite{Kodama:2011vs}. A more
accurate numerical study will be subject of future work. The main
conclusions, that we presently achieve, are as follows:
\begin{itemize}
\item
Inflation proceeds in the waterfall regime, provided
$\kappa\ll M^2/{\Mpl}^2$ or $\kappa\ll m_{\rm FI}^2/{\Mpl}^2$,
respectively. More accurate relations are given by Eqs.~(\ref{rel:kappa})
and~(\ref{e-folds:ph2}).
\item
When the number of e-folds of inflation in the waterfall regime
is much larger than $60$, the dynamics can be well approximated
as in Section~\ref{sec:coarse}. There is no restriction on
the scale of symmetry breaking, but the normalization of
the power spectrum imposes a relation with the superpotential
coupling $\kappa$, {\it cf.} Eqs.~(\ref{P:R},~\ref{P:R:ph1}) and FIGs.~\ref{fig:FtermCMB},~\ref{fig:DtermCMB},~\ref{fig:DtermCMB:2D}.
The spectral index $n_{\rm s}$, Eq.~(\ref{ns:coarse}) ({\it cf.}
also Ref.~\cite{Kodama:2011vs}), then takes values below its
present central observational value.  It can be in agreement (but in strong tension) with WMAP only for the D-term model with $0.1 \Mpl \lesssim m_{\rr{FI}} \lesssim \Mpl$, $g\sim \mathcal O(1)$ and $\kappa \approx 10^{-4}$. 
\item
It is therefore interesting to study how larger values of $n_{\rm s}$
can be achieved. In Section~\ref{section:refined}, we have shown that
provided the largest observed scales leave the horizon close to
the critical point, the deviation of $n_{\rm s}$ from unity
is suppressed as $M^2/{\Mpl}^2$ or $m_{\rm FI}^2/{\Mpl}^2$,
respectively. The point, where just enough e-folds of inflation
in the waterfall regime occur and the power spectrum is normalized
in accordance with the observed values can be inferred from
FIGs.~\ref{fig:FtermCMB},~\ref{fig:DtermCMB},~\ref{fig:DtermCMB:2D}.
In particular, for $F$-term inflation, we can estimate that
this situation occurs for $\kappa\approx 10^{-12}$ and
$M\approx 5\times 10^{12}\,{\rm GeV}$.
\end{itemize}

When the observed limits on $n_{\rm s}$ further tighten around its 
presently observed value, it is in order to further study the latter
possibility. For the purpose of determining the parameters more
accurately than
by the order-of-magnitude estimate in the present work, it
will be necessary to perform a numerical study instead of the
present analytic approximations. In particular, this is necessary
because of the inaccuracies in the matching between the phases~1
and~2, as explained in Section~\ref{section:refined}, and because we did not consider 
the possible contribution of iso-curvature modes.
An interesting
question will be how the observational uncertainty in $n_{\rm s}$
will translate into allowed ranges of $\kappa$ and $M$ or $m_{\rm FI}$,
because this will quantify the amount of parametric tuning that is
required for SUSY hybrid inflation to match observations.

Finally, we have mentioned for the D-term model an extreme case in which the energy scale of inflation is only a few TeV, so that the model can provide a mechanism to the current cosmic acceleration of the expansion, accordingly to Ref.~\cite{Ringeval:2010hf}.  This occurs for $g\approx 10^{-20}$, $\kappa \approx 10^{-16} $ and $m_{\rr{FI}} \approx 10^{-8} \Mpl$.

We conclude that SUSY hybrid models will remain interesting proposals
in order to explain the observations of the primordial power spectrum,
even when it is further confirmed that $n_{\rm s}$ is substantially
below $0.98$. Even without effects from additional non-renormalizable
operators, smaller values of $n_{\rm s}$ are predicted provided
the largest observed scales left the horizon during the waterfall 
regime. Note that also in this situation, the appealing features of the model, which are
the dependence on a small number (2 or 3) of renormalizable operators
only, the natural emergence of inflationary field configurations
due to the attractor property of the potential and finally, the
motivation from SUSY~\cite{Dine:2011ws} remain intact.

\subsubsection*{Acknowledgements}

\noindent
 It is a pleasure to thank Anne-Christine Davis, Jer\^ome Martin, Christophe Ringeval and Vincent Vennin for fruitful discussions and comments.  
The work of BG is supported by the Gottfried Wilhelm Leibniz programme
of the Deutsche Forschungsgemeinschaft.  S.C. is supported by the Wiener-Anspach foundation. 

\bibliography{biblio}

\end{document}